\documentstyle[aps,prl]{revtex} 
\input epsf.sty 

\begin{document}
\renewcommand{\theequation}{\arabic{equation}} 

\twocolumn[ 
\hsize\textwidth\columnwidth\hsize\csname@twocolumnfalse\endcsname 
\draft 

%%%%%%%%%%%%%%%%%%%%%%%%%%%%%%%%%%%%%%%%%%%%%%%%
\title{Ginzburg-Landau Theory for Unconventional Superconductors: \\
Noncompact U(1) Lattice Gauge Model Coupled with Link Higgs Field 
}

\author{Tomoyoshi Ono and Ikuo Ichinose} 
\address{Department of Applied Physics,
Nagoya Institute of Technology, Nagoya, 466-8555 Japan 
}

\date{\today}

\maketitle 

\begin{abstract}   
In this paper, we introduce a Ginzburg-Landau (GL) theory for the extended-$s$
and $d$-wave superconductors (SC) in granular systems that is defined
on a lattice.
In contrast to the ordinary Abelian Higgs model (AHM) that is a 
GL theory for the 
$s$-wave SC, Cooper-pair field (Higgs field) is put on links of the 
lattice in the present model.
By means of Monte-Carlo (MC) simulations, we study phase structure,
gauge-boson mass (the inverse magnetic penetration depth) and 
density of instantons.
In the ordinary {\em noncomapct} U(1) AHM, there exists a second-order phase
transition from the normal to SC states and the gauge-boson mass develops
continuously from the phase transition point.
In the present gauge system with link Higgs field, on the other hand, 
phase transition to the SC state
is of first order at moderate coupling constants.
The gauge-boson mass changes from vanishing to finite values discontinuously
at the phase transition points.

\end{abstract} 
]%the end of twocolumn 

\setcounter{footnote}{0} 
%%%%%%%%%%%%%%%%%%%%%%%%%%%%%%%%%%%%%%%%%%%%%%%%%%%%%%%%%%%%%%%%%

Ginzburg-Landau (GL) theory plays a very important role for study on
superconducting (SC) phase transition.
By taking into account the effect of the electromagnetic interactions,
one can show that
it has a form of a noncompact U(1) lattice-gauge-Higgs model
in which the order-parameter boson
field sits on lattice sites\cite{sWGL}. 
In the last few decades, unconventional superconductors, whose
order parameter of Bose condensation is not the usual $s$-wave, 
have been discovered\cite{UCSC}.
In this paper we shall extend the GL theory to that for unconventional
SC like the $d_{x^2-y^2}$-wave SC in which the order-parameter field,
the Cooper-pair wave function, changes its sign under a $\pi/2$
rotation of the real-space coordinates.
Therefore in order to describe such an unconventional SC, the order-parameter
field, Higgs boson field, must be put on lattice links instead of 
lattice sites.
On-site amplitude of the Cooper pair is vanishingly small because
of, e.g., the strong on-site Coulomb repulsion.

We shall use the path-integral formalism, and define
the model on a three-dimensional (3D) cubic lattice of
system size $L^3$ with the periodic boundary condition.
Before going into details of the extended model, let us first
consider the ordinary {\em noncompact} Abelian Higgs model (AHM)
as a reference system, which is defined by the following action,
\begin{equation}
A_{\rm AHM}={1\over 2}\Big[\sum_{\rm pl}c_u F^2_{ij}(x)
  +\sum_{\rm link}\kappa \;\phi_{x+j}U_{x,j}\phi^\dagger_x\Big],
\label{AHM}
\end{equation}
where $F_{ij}(x)=A_{x,i}-A_{x+j,i}+A_{x+i,j}-A_{x,j}\; (i,j=1,2,3)$ 
and a gauge field $A_{x,i}$ on the link $(x,i)$ is related to
the electromagnetic vector potential $\vec{A}^{\rm em}_i$ as
$A_{x,i}=\int^{x+i}_x \vec{A}^{\rm em}\cdot d\vec{\ell}$ and 
$U_{x,j}=e^{iA_{x,j}}$.
$\phi_x$ is the Higgs field corresponding to the $s$-wave Cooper pair
and in the London limit $\phi_x=e^{i\varphi_x}\; (\varphi_x\in [-\pi,\pi])$.
$c_u^{-1}$ is the electric charge of the Cooper pair and
$\kappa$ is a parameter corresponding to the superfluid density and 
a decreasing function of the temperature ($T$).
$A_{\rm AHM}$ is nothing but the 3D XY model coupled with the noncompact
U(1) gauge field that describes the electromagnetic interactions.

\begin{figure} 
  \begin{picture}(0,80) 
    \put(-5,0){\epsfxsize 120pt  
    \epsfbox{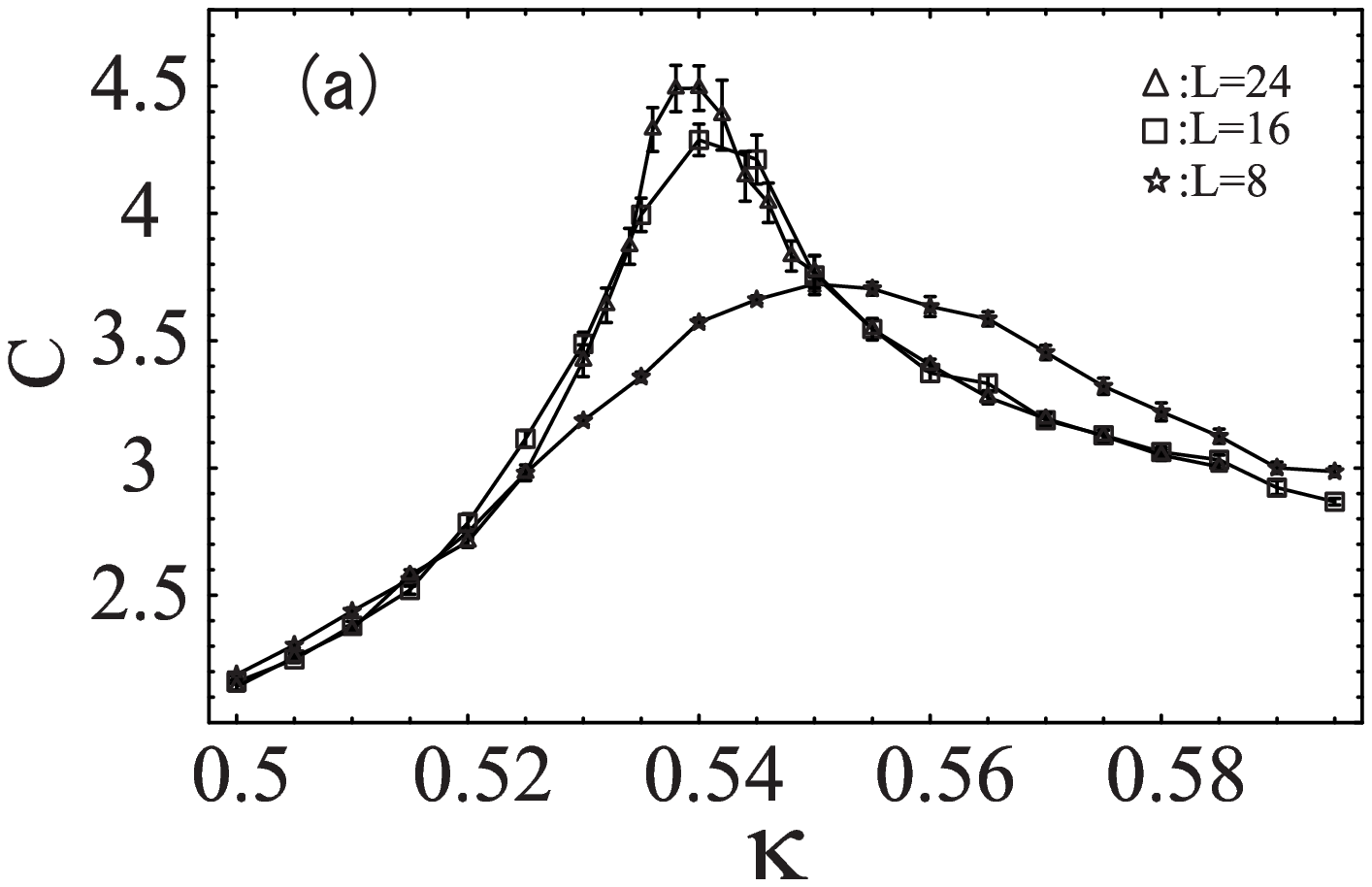}} 
    \put(122,3){\epsfxsize 120pt  
    \epsfbox{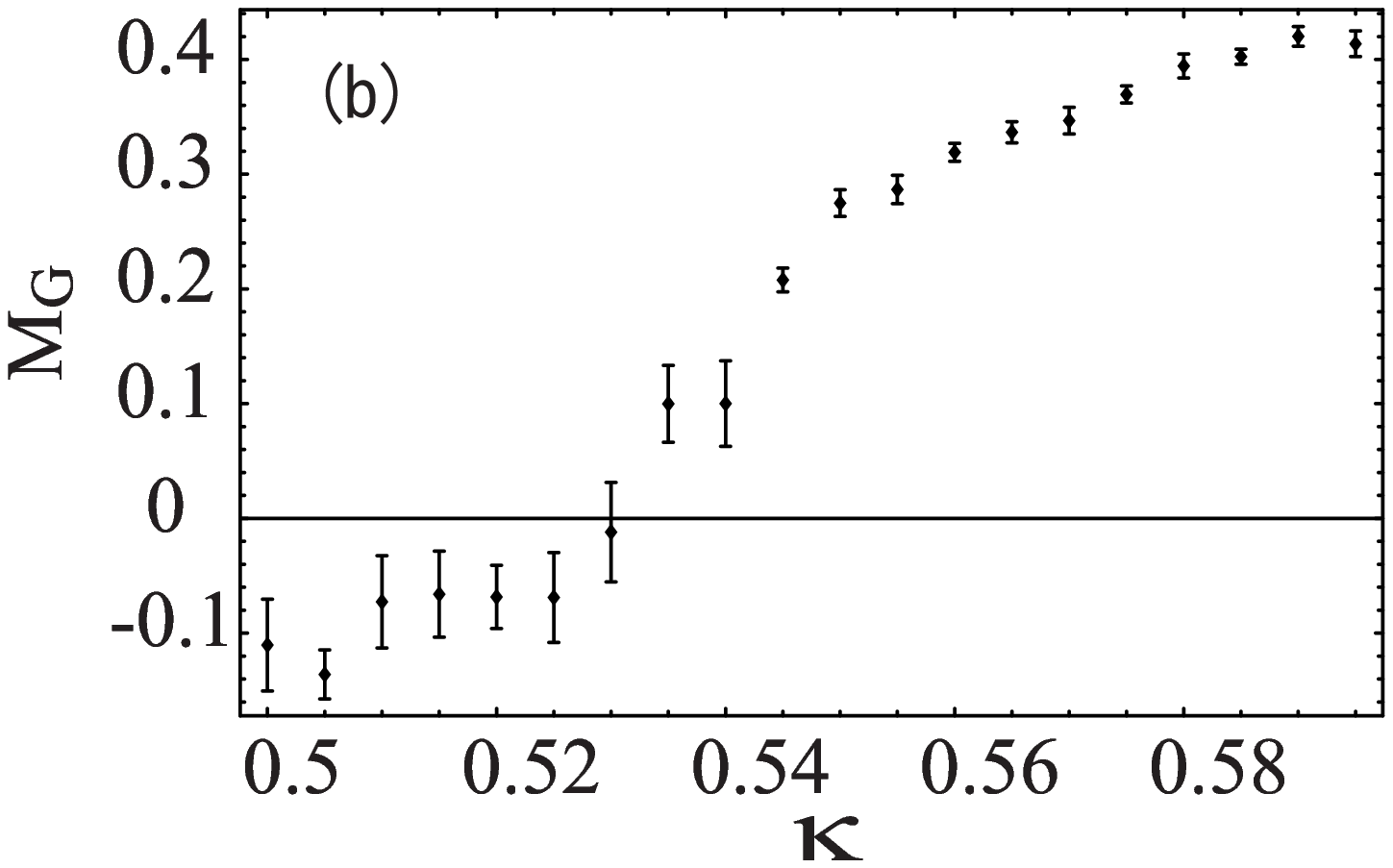}} 
  \end{picture}
\vspace{0.3cm}
\caption{Results of the MC simulations for the AHM with $c_u=1$.
(a) Specific heat $C$ as a function of $\kappa$.
It indicates the existence of a second-order phase transition at 
$\kappa_c=0.538\; (L=24)$.
(b) Gauge-boson mass $M_G$ obtained from the correlation function of
$\sin (F_{ij}(x))$. The critical coupling is estimated as $\kappa_c=0.52 \;
(L=16)$.
}
\label{AHM1}
\end{figure}

We studied the phase structure of the model $A_{\rm AHM}$ by the 
Monte-Carlo (MC) simulations
calculating the internal energy $E=-\langle A_{\rm AHM} \rangle/L^3$
and the specific heat $C=\langle (A_{\rm AHM}-\langle A_{\rm AHM}\rangle
)^2\rangle/L^3$.
We found that there exists a second-order
phase transition line\cite{AHM} emanating from the 3D XY critical point
at $(\kappa,c_u)=(0.46, \infty)$ (obtained in the system of size 
$L=24$).
In Fig.\ref{AHM1}(a), we show the specific heat as a function of 
$\kappa$ with $c_u=1$.
The specific heat exhibits a typical behavior of the second-order phase 
transition as the system size $L^3$ is increased from $8^3$ to $24^3$.
This result is in sharp contrast to the {\em compact} AHM in the 
London limit in which no phase transitions occur and 
only the confinement phase exists\cite{onephase}.

We also measured the gauge-invariant gauge-boson mass $M_G$ which is
defined through the correlation function of the operator $\sin (F_{ij}(x))$.
More precisely let us define the operator $O(x)$ as,
\begin{equation}
O(x)=\sum_{i,j=1,2}\epsilon_{ij}\sin F_{ij}(x),\;\; 
\epsilon_{12}=-\epsilon_{21}=1,
\label{O}
\end{equation}
and its Fourier transformed operator in the $1-2$ plane,
\begin{equation}
\tilde{O}(x_3)=\sum_{x_1,x_2}O(x)e^{ip_1x_1+ip_2x_2}.
\label{tildeO}
\end{equation}
One can expect that the correlation function of $\tilde{O}(x_3)$ behaves as
\begin{equation}
\langle \tilde{O}(x_3)\tilde{O}(x_3+t)\rangle \propto
e^{-\sqrt{p_1^2+p_2^2+M^2_G}\; t}.
\label{OO}
\end{equation}
In Fig.\ref{AHM1}(b), we show the result.
We define the gauge-boson mass $M_G$ from numerical results as 
$M_G={\rm sign}\;(\lambda^2 - \vec{p}^2)\sqrt{\lambda^2-\vec{p}^2}$, 
where $\lambda$ is the inverse
correlation length of the Fourier transformed operator $\tilde{O}(x_3)$
with finite momentum $\vec{p}$ \cite{Mass}.
The negative values for $\kappa< 0.52$ in Fig.\ref{AHM1}(b) come from 
the above definition of the mass $M_G$ and it is considered as 
a finite-size effect\cite{Mass}.

From Fig.\ref{AHM1}(b),
we conclude that $M_G$ is vanishing for $\kappa< 0.52$ and develops
continuously as $\kappa$ increases.
This behavior obviously is consistent with the specific heat measurement
in Fig.\ref{AHM1}(a) and indicates the existence of a second-order phase 
transition from the normal to the Higgs-SC phases, though the value of
the critical coupling $\kappa_c$ obtained from $M_G$ 
($\kappa_c=0.52$ with $L=16$) 
is slightly different from that obtained by $C$ ($\kappa_c=0.54$ with $L=16$).
Similar phenomenon has been observed also in the previous studies on 
the U(1) gauge field coupled with the CP$^1$ fields\cite{Mass2}.

%---------------------------------------------------

%---------------------------------------------------

We applied the finite-size scaling to $C$ as 
$C(\kappa, L)=L^{\sigma/\nu}\phi(L^{1/\nu}\epsilon)$, 
$\epsilon=(\kappa-\kappa_\infty)/\kappa_\infty$, where $\kappa_\infty$
is the critical coupling at $L\rightarrow \infty$ and
$\phi(x)$ is the scaling function.
We found that $\kappa_\infty=0.534, \; \nu=0.83$ and $\sigma=0.15$.

Let us introduce the extended mode that is defined by
the following action,
\begin{eqnarray}
A_{\rm GL}&=&{1\over 2}\sum_{\rm pl}\Big[c_u F^2_{ij}(x)
+c_vV^4+c_m(UVUV+VUVU)  \nonumber  \\
&& +d_m(UUVV+\mbox{3 cyclic permutations})\Big],
\label{AGL}
\end{eqnarray}
where $V_{x,j}$ is the spin-singlet Cooper-pair field on link
that is related to electron operator
$\psi_{x\sigma}\;(\sigma=\uparrow, \downarrow)$ as
\begin{equation}
V_{x,j}\propto \langle \psi_{x\uparrow}\psi_{x+j\downarrow}
-\psi_{x\downarrow}\psi_{x+j\uparrow}\rangle.
\label{V}
\end{equation}
From Eq.(\ref{V}), it is obvious that $V_{x,j}$ can be regarded as a 
three-component vector field.
Here it is interesting to notice that a GL theory for the spin-triplet
$p$-wave superconductivity in ferromagnetic ZrZn$_2$ was proposed
and it employs a SC order parameter similar to $V_{x,j}$\cite{ZrZn}.
Gradient terms of the GL theory for ZrZn$_2$ have a similar form to $A_{\rm GL}$in Eq.(\ref{AGL}).

\begin{figure} 
  \begin{picture}(0,105) 
    \put(30,0){\epsfxsize 180pt  
    \epsfbox{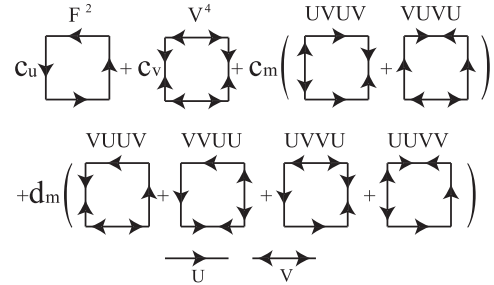}} 
  \end{picture}
  \caption{Action $A_{\rm GL}$ of the GL theory (\ref{AGL}). 
} 
\label{figAGL}
\end{figure}

Hereafter we shall consider the London limit of $V_{x,j}$ and 
set $V_{x,j}=e^{i\theta_{xj}}\; (\theta_{xj}\in [-\pi,\pi])$.
Each term in $A_{\rm GL}$ is depicted in Fig.\ref{figAGL}
where $c_u, \; c_v$ etc are coupling constants, and 
$A_{\rm GL}$ is constructed to be invariant under the following 
{\em noncompact} local gauge transformation,
\begin{equation}
A_{x,j}\rightarrow A_{x,j}+\alpha_{x+j}-\alpha_x,
\;\; V_{x,j}\rightarrow e^{i\alpha_{x+j}}V_{x,j}e^{i\alpha_x}.
\label{gaugetr}
\end{equation}
From Eq.(\ref{gaugetr}), it is obvious that $V_{x,j}$ can be regarded as 
another gauge field dual to the electromagnetic gauge field\cite{FN1}.
We consider terms as local as possible for $A_{\rm GL}$,
and the partition function $Z$ is given as 
\begin{equation}
Z=\int_{-\infty}^{\infty}[DA]\int_{-\pi}^{\pi}[D\theta]\;
e^{A_{\rm GL}}.
\label{Z}
\end{equation}

%---------------------------------------------------
%----------------------------------------------------

Compact U(1) version of the above $A_{\rm GL}$, in which 
$\sum_{\rm pl}F^2_{ij}(x)$ is
replaced by $\sum_{\rm pl}U^4$, has been studied in the
previous paper\cite{UV1}.
In the present paper, we study the noncompact U(1) gauge theory as a
GL theory for the unconventional SC in which the gauge field $U_{x,j}$
describes the electromagnetic field.

There is credible evidence that the SC phase transition in the high-$T_c$
cuprates is of second-order and furthermore it is in the 3D XY model 
universality class\cite{XY}.
On the other hand
as explained above, the Cooper-pair field must sit on lattice links
instead of sites in order to describe the $d$-wave SC state.
One of the simplest GL theory for the $d$-wave SC is $A_{\rm GL}$ given in 
(\ref{AGL}).
In the present paper, we shall study the phase structure and 
physical properties of $A_{\rm GL}$
by means of the MC simulations and compare the results with those of the 
noncompact AHM and the XY model.

Let us first notice that
for vanishing $c_m=d_m=0$, the present system reduces to 
two independent decoupled gauge models, a noncompact U(1) gauge model of 
$A_{x,j}$ and a compact U(1) gauge model of $V_{x,j}$.
Then it is obvious that there exist no phase transitions in that case.
By giving finite values for $c_m$ and/or $d_m$, we study 
phase structure in the $c_u-c_v$ plane.

We first study the case $c_m=1,\; d_m=0$.
It is instructive to consider the large-$c_u$ limit in which configurations
of the noncompact gauge field are restricted as $A_{x,j}\sim
\varphi_{x+j}-\varphi_x$.
Then the $c_m$-terms in $A_{\rm GL}$ become as 
\begin{equation}
\sum_{x,i,j}e^{-i\varphi_x}V_{x,i}e^{-i\varphi_{x+i}}\cdot
e^{i\varphi_{x+j}}V^\dagger_{x+j,i}e^{i\varphi_{x+i+j}}
+\mbox{c.c.}
\label{cm1}
\end{equation}
%Comparing with the Higgs coupling in Eq.(\ref{AHM}) (the $\kappa$-term), 
We shall call the above term double Higgs coupling.
As we explained above, the usual Higgs coupling of the compact gauge field
$V_{x,j}$,
$\sum_{x,j}e^{-i\varphi_x}V_{x,j}e^{-i\varphi_{x+j}}$, does {\em not} induce
any phase transition.
On the other hand, the doubly-charged Higgs coupling,
$e^{-i\varphi_x}(V_{x,i})^2e^{-i\varphi_{x+i}}$,
induces a phase
transition from the confinement to Higgs phases\cite{AHM2}. 
Then it is interesting to study the extended model 
also from the viewpoint of the Higgs coupling
and see if the Higgs phase 
transition occurs as the double Higgs coupling (\ref{cm1}) is increased.

\begin{figure} 
  \begin{picture}(0,80) 
    \put(0,0){\epsfxsize 110pt  
    \epsfbox{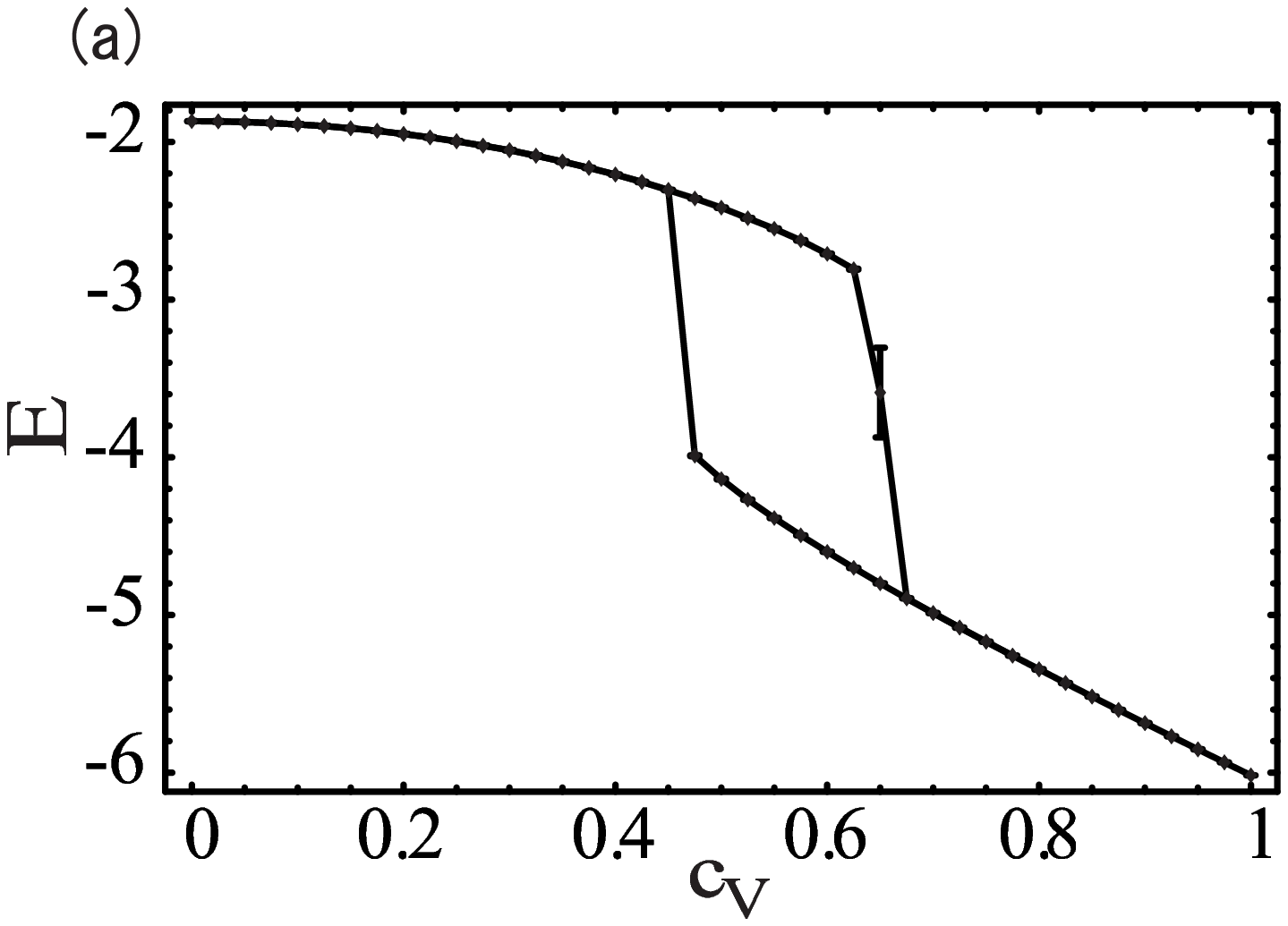}} 
    \put(120,0){\epsfxsize 110pt  
    \epsfbox{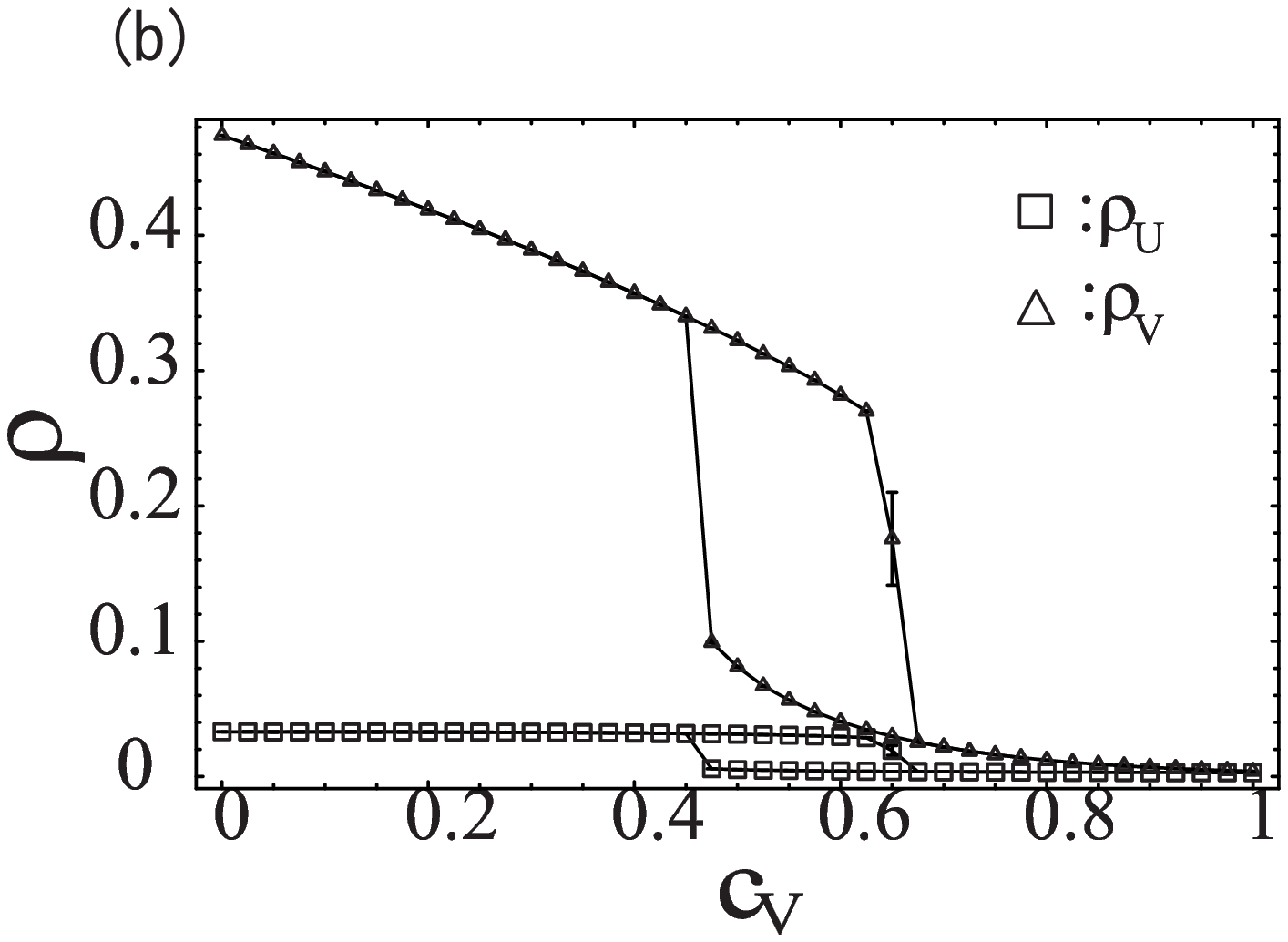}} 
  \end{picture}
\vspace{0.3cm}
\caption{Results for $A_{\rm GL}$ with $c_m=1,\; d_m=0,\; c_u=0.5$ ($L=16$).
(a) Internal energy exhibiting hysteresis loop.
(b) Densities of instanton. $\rho_U$ ($\rho_V$) is the instanton
density of the noncompact gauge field $A_{x,j}$ (compact gauge
field $V_{x,j}$).
}
\label{Es1}
\end{figure}

We studied the phase structure in the $c_u-c_v$ plane by 
calculating $E$ and $C$ and found that there is a phase transition line.
Typical behavior of $E$ near phase transition points 
is shown in Fig.\ref{Es1}(a),
which indicates that the transition is of first-order. 
To see physical meaning of the phase transition, we measured the instanton
densities of the gauge fields $U_{x,j}$ and $V_{x,j}$.
We follow the definition of the instanton densities $\rho_U$ and $\rho_V$
given in Refs.\cite{instanton,UV1}.
The result is shown in Fig.\ref{Es1}(b).
As $U_{x,j}$ is the noncompact gauge field, the density of instanton 
$\rho_U$ is vanishingly small.
On the other hand for $V_{x,j}$, $\rho_V$ exhibits a hysteresis loop 
just like the internal energy $E$ at the critical point.
Vanishing of $\rho_V$ for 
$c_v>0.4(0.6)$ means that the observed phase transition is the normal to 
Higgs-SC phase transition.
As in the compact U(1) gauge case\cite{UV1}, adding small but finite positive
$d_m$-terms stabilizes the sign of $\langle UUVV \rangle$ as 
$\langle UUVV \rangle >0$ in the Higgs-SC phase.
This SC phase corresponds to the extended $s$-wave, because on-site
amplitude of the Cooper pair is zero whereas {\em expectation values} 
of $V_{x,j}$ 
on links $(x,j)\; (j=1,2,3)$ have the same sign under the gauge-fixing
condition $\varphi_x=0$ in Eq.(\ref{cm1})\cite{QF}.

%---------------------------------------------------

%---------------------------------------------------
%---------------------------------------------------
\begin{figure} 
  \begin{picture}(0,70) 
    \put(30,0){\epsfxsize 140pt  
    \epsfbox{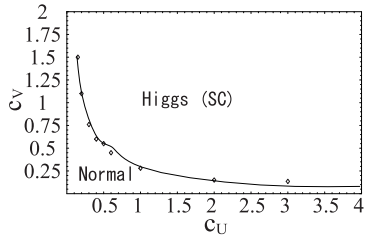}} 
  \end{picture}
  \caption{Phase diagram of $A_{\rm GL}$ with $c_m=1,\; d_m=0$.
Locations of the phase transition points are determined
by those of center of the hysteresis loops.
``Normal" phase denotes
the Coulomb (confinement) phase of $U_{x,j}$ ($V_{x,j}$).
} 
\label{Espd}
\end{figure}
%----------------------------------------------------

%---------------------------------------------------
\begin{figure} 
  \begin{picture}(0,80) 
    \put(-10,0){\epsfxsize 125pt  
    \epsfbox{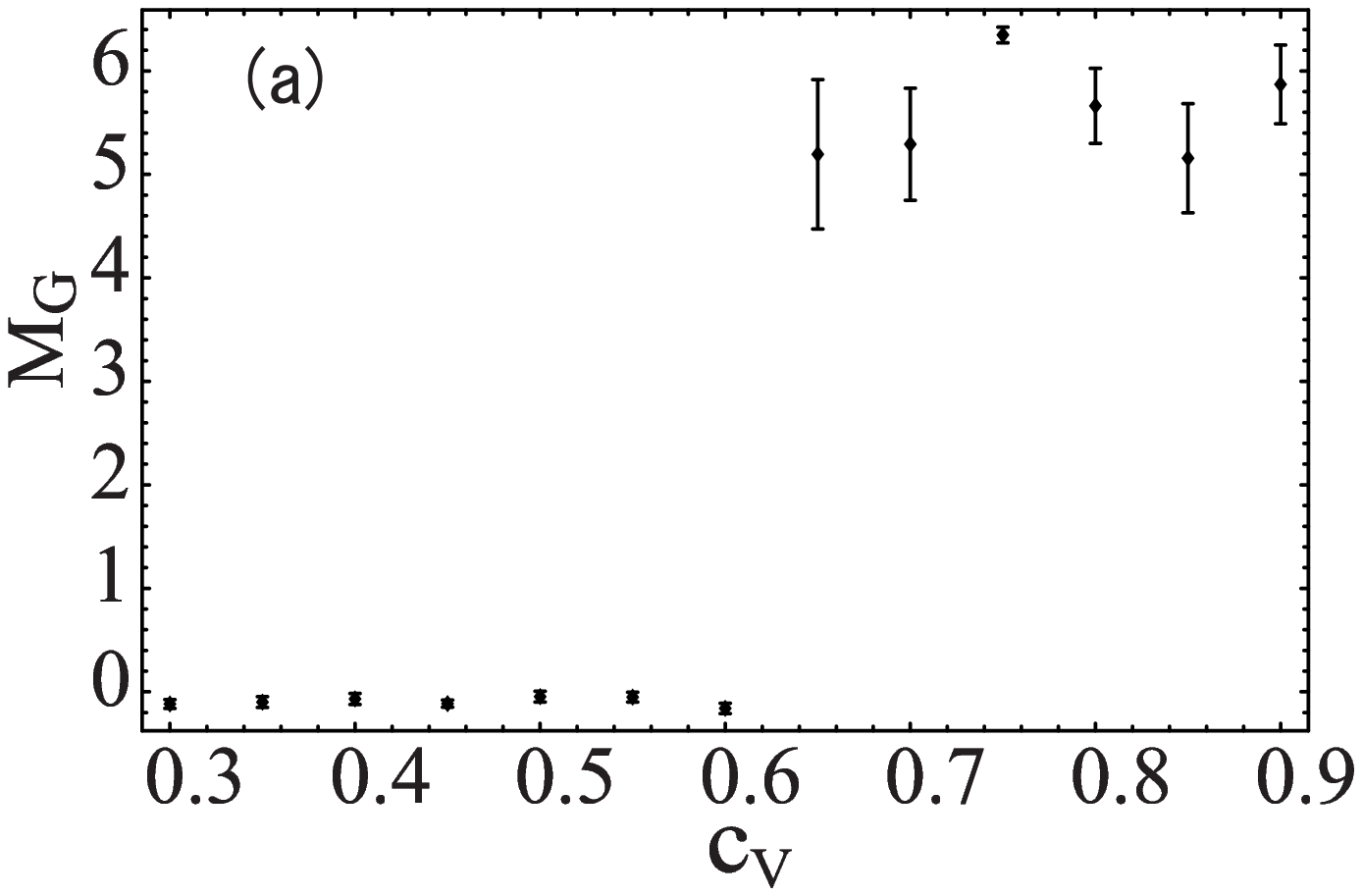}} 
    \put(120,0){\epsfxsize 120pt  
    \epsfbox{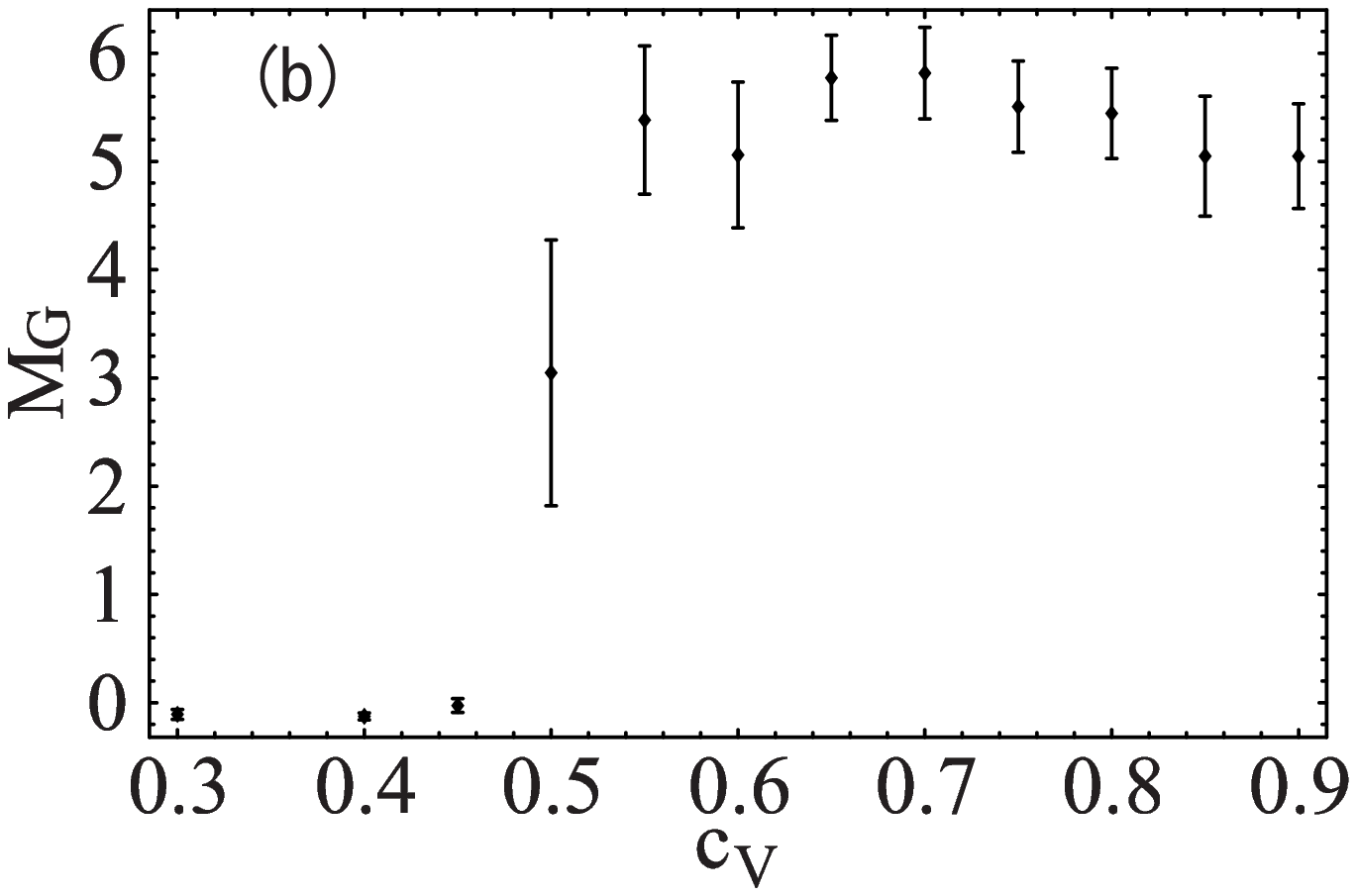}} 
  \end{picture}
\vspace{0.3cm}
\caption{Results for $A_{\rm GL}$ with $c_m=1,\; d_m=0, \; c_u=0.5$.
(a) Gauge-boson mass ($L=16$) measured for increasing $c_v$. 
It exhibits a sharp discontinuity at $c_v=0.6$, the phase transition point.
(b)Gauge-boson mass for decreasing $c_v$. 
}
\label{Es2}
\end{figure}

%---------------------------------------------------

In Fig.\ref{Espd}, we show the phase diagram obtained from the measurement
of the internal energy.
In Fig.\ref{Es2}(a) and (b), calculations of the mass of the gauge boson 
$A_{x,j}$ are given.
Contrary to the AHM in Fig.\ref{AHM1}, the gauge-boson mass exhibits a sharp
discontinuity and acquires nonvanishing value at $c_v=0.60(0.45)$.

\begin{figure} 
  \begin{picture}(0,90) 
    \put(30,0){\epsfxsize 120pt  
    \epsfbox{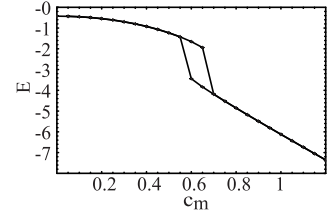}} 
  \end{picture}
  \caption{$E$ as a function of $c_m$ for $c_u=c_v=1$.
} 
\label{figc_m}
\end{figure}

We also studied phase transition with $c_u,\; c_v$ fixed and $c_m$ varied,
and found that a first-order phase transition occurs at a critical
coupling of $c_m$.
Measurement of $E$ is given in Fig.\ref{figc_m} for the $c_u=c_v=1$ case.
Density of instantons exhibits similar behavior.

%---------------------------------------------------
%----------------------------------------------------

Let us turn to the $d_m< 0$ case.
In the large-$c_u$ limit, $U_{x,j}\sim e^{i\varphi_{x+j}}e^{-i\varphi_x}$
and the $d_m$-terms in $A_{\rm GL}$ prefer configurations like 
$\langle e^{-i\varphi_{x+i}}V_{x,i}V^\dagger_{x,j}e^{i\varphi_{x+j}}
\rangle<0\;(i\neq j)$.
Then under the gauge-fixing condition $\varphi_x=0$, it is expected that
the expectation value of the Cooper-pair field
$\langle V_{x,i}\rangle$ changes its sign under a $\pi/2$
rotation in a plane.
Though some of the $d$-wave SC materials have a layered structure,
we first consider the 3D isotropic case and set $d_m=-0.8$.
%For values of $0>d_m>-0.8$, we did not find the phase transitions.

\begin{figure} 
  \begin{picture}(0,80) 
    \put(-5,-2){\epsfxsize 125pt  
    \epsfbox{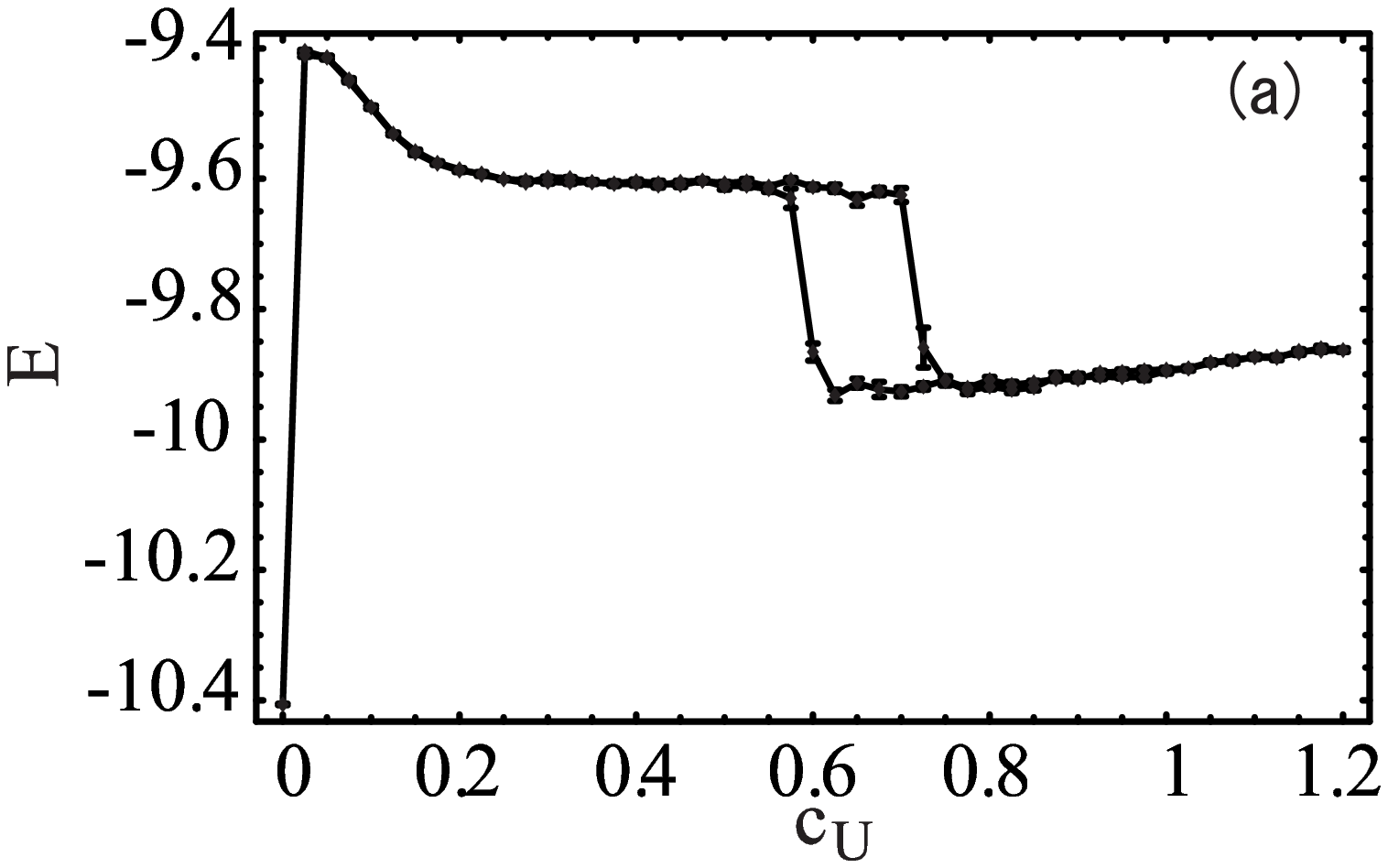}} 
    \put(120,0){\epsfxsize 115pt  
    \epsfbox{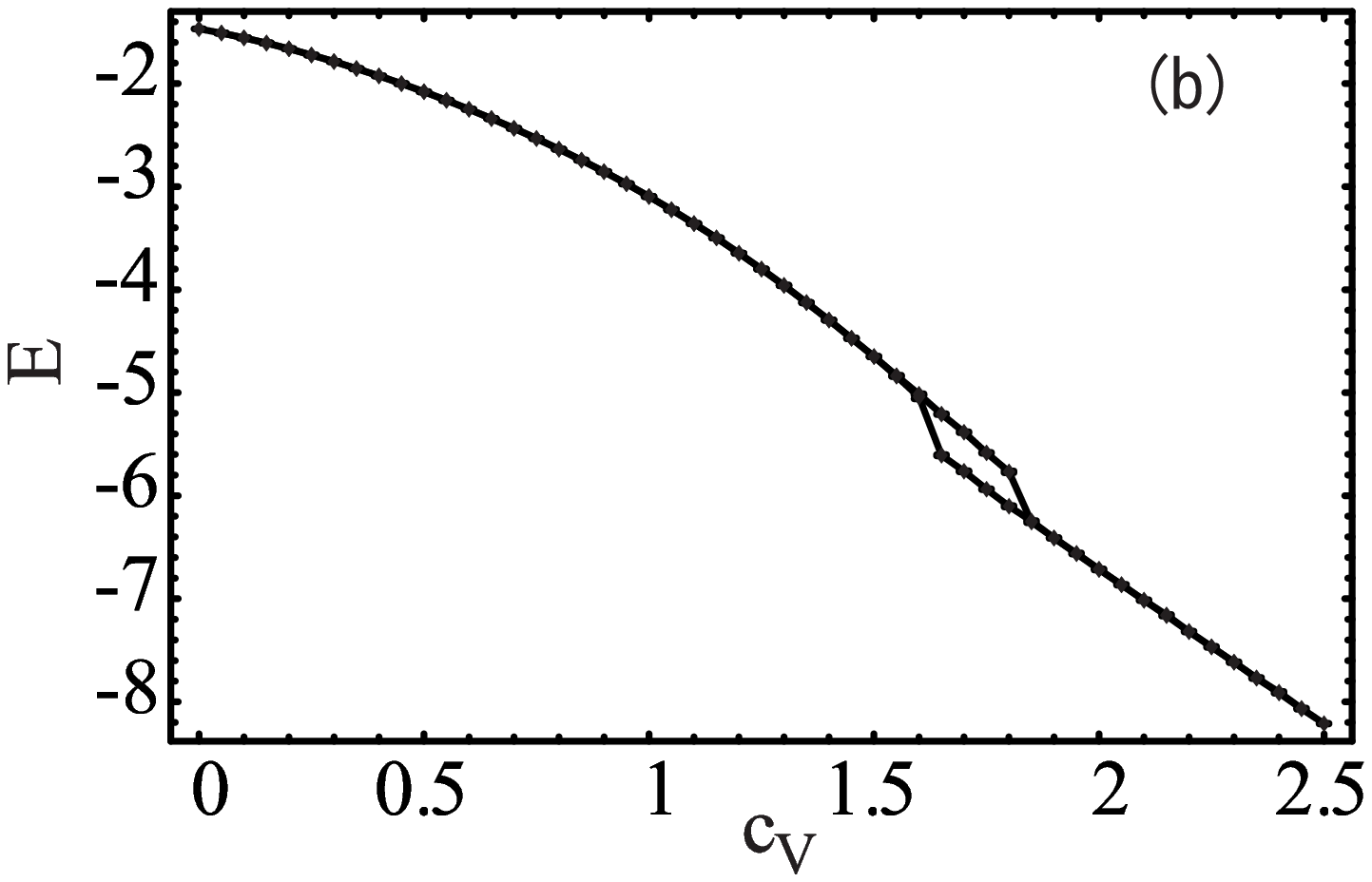}} 
  \end{picture}
\vspace{0.3cm}
\caption{Results for $A_{\rm GL}$ with $c_m=0,\; d_m=-0.8$.
(a) $E$ as a function of $c_u$ for $c_v=3$.
(b) $E$ as a function of $c_v$ for $c_u=3$.
}
\label{d1}
\end{figure}

\begin{figure} 
  \begin{picture}(0,70) 
    \put(30,0){\epsfxsize 140pt  
    \epsfbox{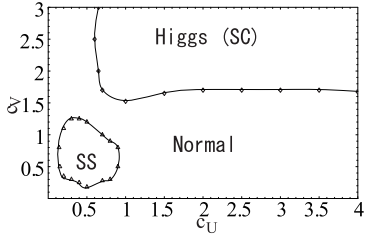}} 
  \end{picture}
  \caption{Phase diagram for $c_m=0,\; d_m=-0.8$.
Locations of the phase transition points are determined
by those of center of the hysteresis loops.
} 
\label{d2}
\end{figure}

Phase structure was studied by means of the MC simulations as before and 
found that there exist phase transition lines.
The internal energy $E$ shows hysteresis loop at critical points as
in the previous case.
We show the calculations of $E$ for certain places in the $c_u-c_v$ 
plane in Fig.\ref{d1}.
Phase diagram obtained by the measurement of $E$ and $C$ is given in
Fig.\ref{d2}.
The gauge-boson mass $M_G$ exhibits a discontinuity at phase transition
points as in the previous case.

%---------------------------------------------------

%---------------------------------------------------

%---------------------------------------------------
%----------------------------------------------------

Besides the normal and Higgs-SC phases, we found that there exists an
exotic phase that we call staggered state (SS).
Existence of a similar phase has been observed for the compact gauge 
model\cite{UV1}.
It stems from the fact that in 3D there are no configurations that
satisfy $V_{x,i}V^\dagger_{x,j}<0\;(i\neq j)$ for {\em all} $i$ and $j$
simultaneously.
In other words, the $d_m$-terms cause frustrations.
$E$ and the instanton density exhibit the first-order phase transition
at the boundary of the normal and SS states. 
However, we also found that as the system size is getting larger, 
signal of the phase transition at the boundary is getting weaker. 
It is possible that finite region of the SS disappears for 
the infinite system-size limit.

%---------------------------------------------------
\begin{figure} 
  \begin{picture}(0,80) 
    \put(-5,-2){\epsfxsize 115pt  
    \epsfbox{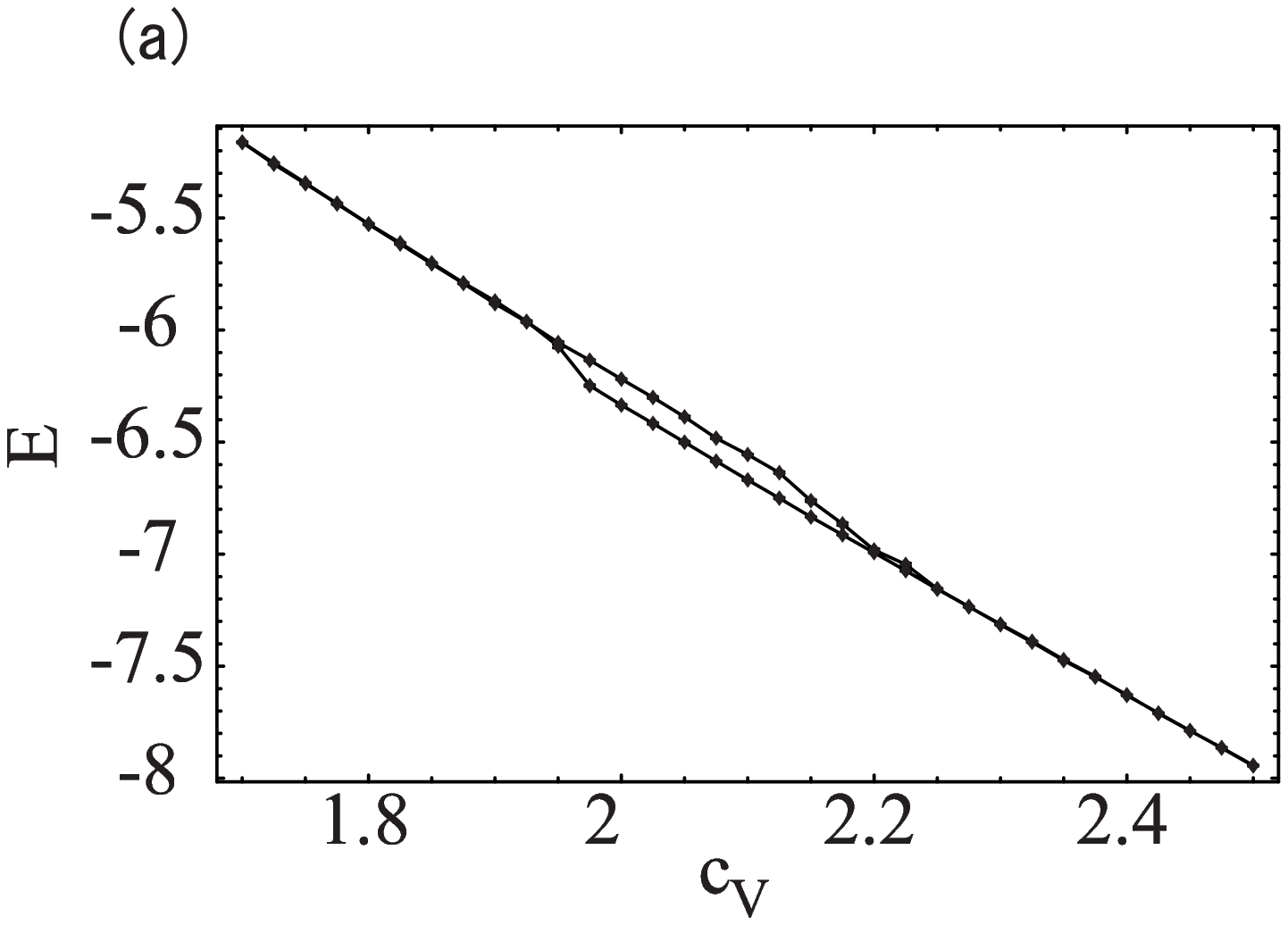}} 
    \put(120,0){\epsfxsize 115pt  
    \epsfbox{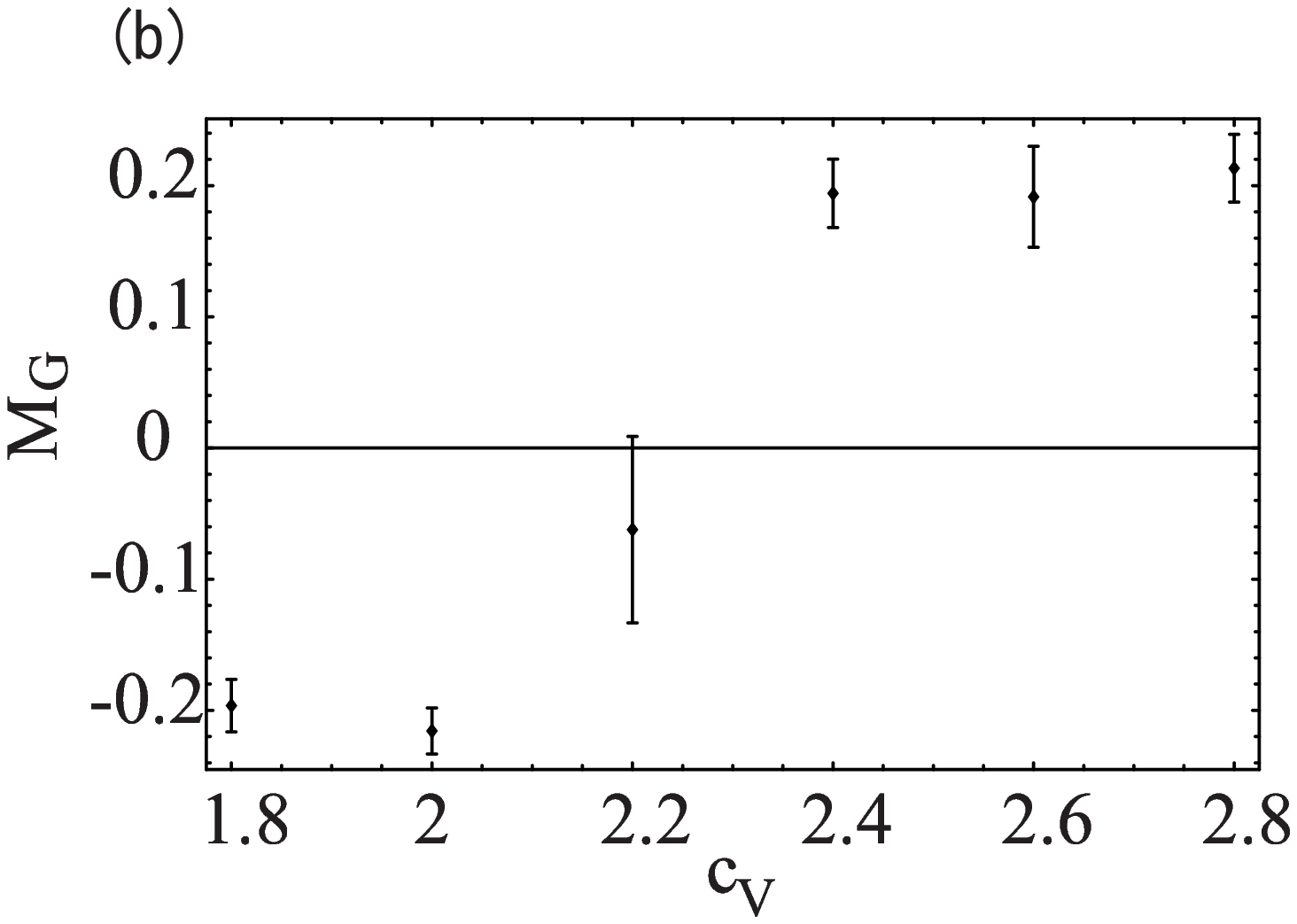}} 
  \end{picture}
\vspace{0.3cm}
\caption{Results for $A_{\rm GL}$ with $c_m=0,\; d_m=-0.8(-0.4)$
for intralayer(interlayer) coupling and $c_u=0.5$ ($L=24$).
(a) $E$ as a function of $c_v$ for $c_u=0.5$.
(b) Measurement of $M_G$ (average of $M_G$ in three directions)
 as a function of $c_v$ for increasing $c_v$.
Values of $M_G$ are much smaller than those in the 3D isotropic case.
}
\label{d3}
\end{figure}

%---------------------------------------------------
Finally let us turn to anistropic cases and study how the 
phase structure changes due to the layered structure.
To this end, we put different values for the interlayer and 
intralayer $d_m$'s.
Numerical results for $d_m=-0.8(-0.4)$ for the intralayer(interlayer) coupling
with $c_u=0.5$ are given in Fig.\ref{d3}.
The SS, which exists in the isotropic case due to the frustration, is not 
observed in this case and the first-order phase transition to the Higgs-SC 
phase is observed instead.
We also verified that critical line of the normal-SC 
phase transition exists as in the isotropic case.
However, discontinuity in the gauge-boson mass $M_G$ at critical points 
becomes smaller than that in the isotropic case.

In conclusion, we studied the noncompact U(1) lattice gauge model 
with link Higgs field that is a GL theory for the unconventional 
SC including the extended-$s$, $d$-wave and also ferromagnetic
$p$-wave SC's.
By means of the MC simulations, we clarified the phase structure.
There exist first-order phase transitions from the normal to 
Higgs-SC phases.
We also observed that as the anisotropy of the layered structure 
is getting larger, 
signal of the first-order phase transition is getting weaker.

%%%%%%%%%%%%%%%%%%%%%%%%%%%%%%%%%%%%%%%%%%%%%%%%%%%%%%%%%%%%%%%%%%


\begin{references}
\bibitem{FN1}In order to see this fact clearly, let us replace $\alpha_x$ with
$-\alpha_x$ for $x \in$ odd site. Then 
$
V_{x,j}\rightarrow e^{i\alpha_{x+j}}V_{x,j}e^{-i\alpha_x},
$
$x \in$ odd site, and 
$
V^\dagger_{x,j}\rightarrow e^{i\alpha_{x+j}}V^\dagger_{x,j}e^{-i\alpha_x},
$
$x \in$ even site.
\end{references}
\end{document}